\documentclass{article} %PP
\usepackage{a4wide,amssymb,epsf} %PP

\newcommand{\lya}{Ly$\alpha$~}
\newcommand{\la}{\lesssim} %PP
\newcommand{\ga}{\gtrsim} %PP

\begin{document}

\title{Effects of Disks on Gravitational Lensing by Spiral Galaxies}

%JO \author{Matthias Bartelmann}
%JO \affil{Max-Planck-Institut f\"ur Astrophysik, P.O.\ Box 1523,
%JO D--85740 Garching, Germany}
%JO \and
%JO \author{Abraham Loeb}
%JO \affil{Harvard-Smithsonian Center for Astrophysics,
%JO 60 Garden Street, Cambridge, MA 02138, USA}

\author{Matthias Bartelmann\\ %PP
Max-Planck-Institut f\"ur Astrophysik\\ %PP
P.O.\ Box 1523, D--85740 Garching, Germany %PP
\and %PP
Abraham Loeb\\ %PP
Harvard-Smithsonian Center for Astrophysics\\ %PP
60 Garden Street, Cambridge, MA 02138, USA} %PP

\maketitle %PP

\begin{abstract}
Gravitational lensing of a quasar by a spiral galaxy should often be
accompanied by damped \lya absorption and dust extinction due to the
intervening gaseous disk. In nearly edge-on configurations, the
surface mass density of the gas and stars in the disk could by itself
split the quasar image and contribute significantly to the overall
lensing cross section. We calculate the lensing probability of a
disk$+$halo mass model for spiral galaxies, including cosmic evolution
of the lens parameters. A considerable fraction of the lens systems
contains two images with sub-arcsecond separation, straddling a nearly
edge-on disk. Because of that, extinction by dust together with
observational selection effects (involving a minimum separation and a
maximum flux ratio for the lensed images), suppress the detection
efficiency of spiral lenses in optical wavebands by at least an order
of magnitude. The missing lenses could be recovered in radio surveys.
In modifying the statistics of damped \lya absorbers, the effect of
extinction dominates over the magnification bias due to lensing.
\end{abstract}

%JO \keywords{cosmology: theory -- gravitational lensing --- quasars:
%JO absorption lines}

\section{Introduction}

Gravitational lensing by a spiral galaxy occurs when the line-of-sight
to a background quasar passes within a few kpc from the center of the
galactic disk. Since galactic disks are rich in neutral hydrogen (HI),
the quasar spectrum is likely to show a damped \lya absorption trough
at the lens redshift. Therefore, the efficiency of blind searches for
gravitational lensing with sub-arcsecond splitting can be enhanced by
more than an order of magnitude, by selecting a subset of all bright
quasars which show a low-redshift ($z\la 1$) damped \lya absorption
with a high HI column density, $\ga10^{21}\,{\rm cm^{-2}}$ (Bartelmann
\& Loeb 1996). Moreover, multiply imaged quasars could be identified
spectroscopically through their multiple-trough absorption
spectrum. The composite spectrum of a lensed quasar is a superposition
of the spectra received from the different images which intersect the
absorbing disk at different locations, probe different HI column
densities, and hence acquire different widths in their damped \lya
troughs (Loeb 1997).

The magnification bias due to lensing changes the statistics of damped
\lya absorbers (DLAs) in quasar spectra by bringing into view quasars
that would otherwise fall below the detection threshold (Bartelmann \&
Loeb 1996; Smette, Claeskens, \& Surdej 1997). For optical
observations, this effect is counteracted by dust extinction in the
lensing galaxy (Malhotra, Rhoads, \& Turner 1997; Perna, Loeb, \&
Bartelmann 1997). The combination of lensing and dust extinction
results in a net distortion of the HI column density distribution of
damped \lya absorbers.

Since galactic disks are thin, their own surface mass density could
exceed the critical value necessary for image splitting when they
happen to be projected nearly edge-on. In such orientations, a pair of
quasar images will straddle the symmetry axis of the disk rather than
its center. When averaged over all possible disk orientations, the
total lensing cross section of the disk$+$halo mass distribution
should still be roughly the same as that of a halo with the same
spherically-averaged mass profile (Wang \& Turner 1997; Keeton \&
Kochanek 1997). However, because of the new predominant image
configurations around edge-on disks, the effects of dust extinction
and HI absorption are substantially different when lensing by the disk
is included in the calculation.

Previous theoretical studies have either ignored the possibility of
lensing by the disk itself (Bartelmann \& Loeb 1996; Smette et
al. 1997), or else ignored the effects of dust extinction or HI
absorption on the lensing statistics (Maller, Flores, \& Primack 1997;
Wang \& Turner 1997; Keeton \& Kochanek 1997). In addition, these
discussions did not examine the implications of evolution in the
spiral lens population on its overall lensing properties and
statistics. In this paper, we include all of the above
ingredients. Our model evolves the number density of galaxies based on
the Press-Schechter (1974) theory for dark matter halos, combined with
simple evolutionary scaling laws for the disk properties inside these
halos (Mo, Mao, \& White 1997).

In \S2, we describe our model for the mass and HI distributions in
galactic disks, as well as our model for their dust content. We also
summarize the adopted scaling laws for the evolution of the lens
population with redshift. In \S3, we compute the lensing cross
sections, imaging probabilities, and lensing statistics of our model
spiral galaxies. The impact of dust on the lensing statistics, and the
effect of lensing on the statistics of damped \lya absorbers, are
examined in \S4. Finally, \S5 summarizes the main conclusions from
this work.

\section{A Lens Model for Spiral Galaxies}

\subsection{Mass Model and Lensing Properties}

We use a mass model for spiral lenses which was recently suggested by
Keeton \& Kochanek (1997). The model consists of oblate, ellipsoidal,
isothermal building blocks for individual mass components. They have
the axisymmetric density distribution
\begin{equation}
  \rho(R,z) = \frac{v_{\rm c}^2}{4\pi Gq_3}\frac{e}{\sin^{-1}e}
  \frac{1}{R^2 + s^2 + x_3^2/q_3^2}\;,
\label{eq:1}
\end{equation}
where $R$ is the distance from the symmetry axis of the ellipsoid,
$x_3$ is the distance from its mid-plane, $q_3$ is the axis ratio, $s$
is the core radius which softens the mass distribution, $v_{\rm c}$ is
the asymptotic circular velocity, and $e$ is the eccentricity of the
mass distribution,
\begin{equation}
  e = (1-q_3^2)^{1/2}\;.
\label{eq:2}
\end{equation}
Hence, each mass component is described by one parameter quantifying
its dynamical properties, namely the circular velocity $v_{\rm c}$,
and two parameters characterizing its shape, namely the core radius
$s$ and the oblateness $q_3$.

The projection of the three-dimensional density $\rho$ results in the
two-dimensional surface mass density
\begin{equation}
  \Sigma(\vec x) = \frac{\Sigma_{\rm cr}}{2}\,b_{\rm p}\,
  \left[q^2(s^2+x_1^2)+x_2^2\right]^{-1/2}\;.
\label{eq:3}
\end{equation}
Here, $\Sigma_{\rm cr}$ is the critical surface mass density for
lensing,
\begin{equation}
  \Sigma_{\rm cr} = \frac{c^2}{4\pi G}\,
  \frac{D_{\rm s}}{D_{\rm l}D_{\rm ls}}\;,
\label{eq:4}
\end{equation}
where $D_{\rm l}$, $D_{\rm s}$, and $D_{\rm ls}$ are the
angular-diameter distances between the observer and the lens, the
observer and the source, and the lens and the source,
respectively. $\vec x=(x_1,x_2)$ is the position vector on the sky,
and
\begin{equation}
  b_{\rm p} = 2\pi\left(\frac{v_{\rm c}}{c}\right)^2\,
  \frac{D_{\rm l}D_{\rm ls}}{D_{\rm s}}\,
  \frac{e}{\sin^{-1}e} =
  b\,\frac{e}{\sin^{-1}e}\;,
\label{eq:5}
\end{equation}
where $b$ is the Einstein radius of a singular isothermal sphere with
a circular velocity $v_{\rm c}$. Finally, $q$ is the projected axis
ratio
\begin{equation}
  q = (q_3^2\cos^2i+\sin^2i)^{1/2}\;,
\label{eq:6}
\end{equation}
where $i$ is the inclination angle of the disk relative to the
line-of-sight, with $i=0$ for an edge-on orientation.

As shown by Keeton \& Kochanek (1997), the lensing potential of the
surface-mass density (\ref{eq:3}) is
\begin{equation}
  \psi(s,q_3) = x_1\alpha_1 + x_2\alpha_2 -
  b_{\rm p}s\,\ln\left[(\varrho + s)^2+(1-q^2)x_1^2\right]^{1/2}\;,
\label{eq:7}
\end{equation}
with
\begin{eqnarray}
  \varrho &=& \left[q^2(x_1^2+s^2)+x_2^2\right]^{1/2}\;,\nonumber\\
  \alpha_1 &=& \frac{b_{\rm p}}{(1-q^2)^{1/2}}\,\tan^{-1}\left[
  \frac{(1-q^2)^{1/2}x_1}{\varrho + s}\right]\;,\nonumber\\
  \alpha_2 &=& \frac{b_{\rm p}}{(1-q^2)^{1/2}}\,\tanh^{-1}\left[
  \frac{(1-q^2)^{1/2}x_2}{\varrho + q^2s}\right]\;.
\label{eq:8}
\end{eqnarray}

We now combine three of the isothermal oblate ellipsoids to a mass
model consisting of a {\em maximal\/} disk and a surrounding halo. The
total lensing potential then reads
\begin{equation}
  \psi = \psi(s_{\rm d},q_{3{\rm d}}) - \psi(r_{\rm d},q_{3{\rm d}}) +
  \psi(s_{\rm h},1)\;,
\label{eq:9}
\end{equation}
where $s_{\rm d,h}$ are the core radii of disk and halo, respectively,
$q_{3{\rm d}}$ is the disk axis ratio, and $r_{\rm d}$ is the disk
truncation radius, or disk radius for simplicity. The first term is
the potential of a disk with an asymptotically flat rotation curve,
axis ratio $q_{\rm 3d}$, and core radius $s_{\rm d}$. The second term
truncates that disk at radius $r_{\rm d}$. The third term adds the
surrounding spherical halo necessary to maintain a flat rotation curve
beyond the disk truncation radius. This is the maximal {\em truncated
Mestel disk\/}\footnote{For $s_{\rm d}\to0$, the disk becomes a Mestel
(1963) disk.} model introduced by Keeton \& Kochanek (1997). All
lensing properties of the combined model can now be calculated in
terms of the potential $\psi$.

\begin{figure}[ht]
  \centerline{\epsfxsize=0.5\hsize\epsffile{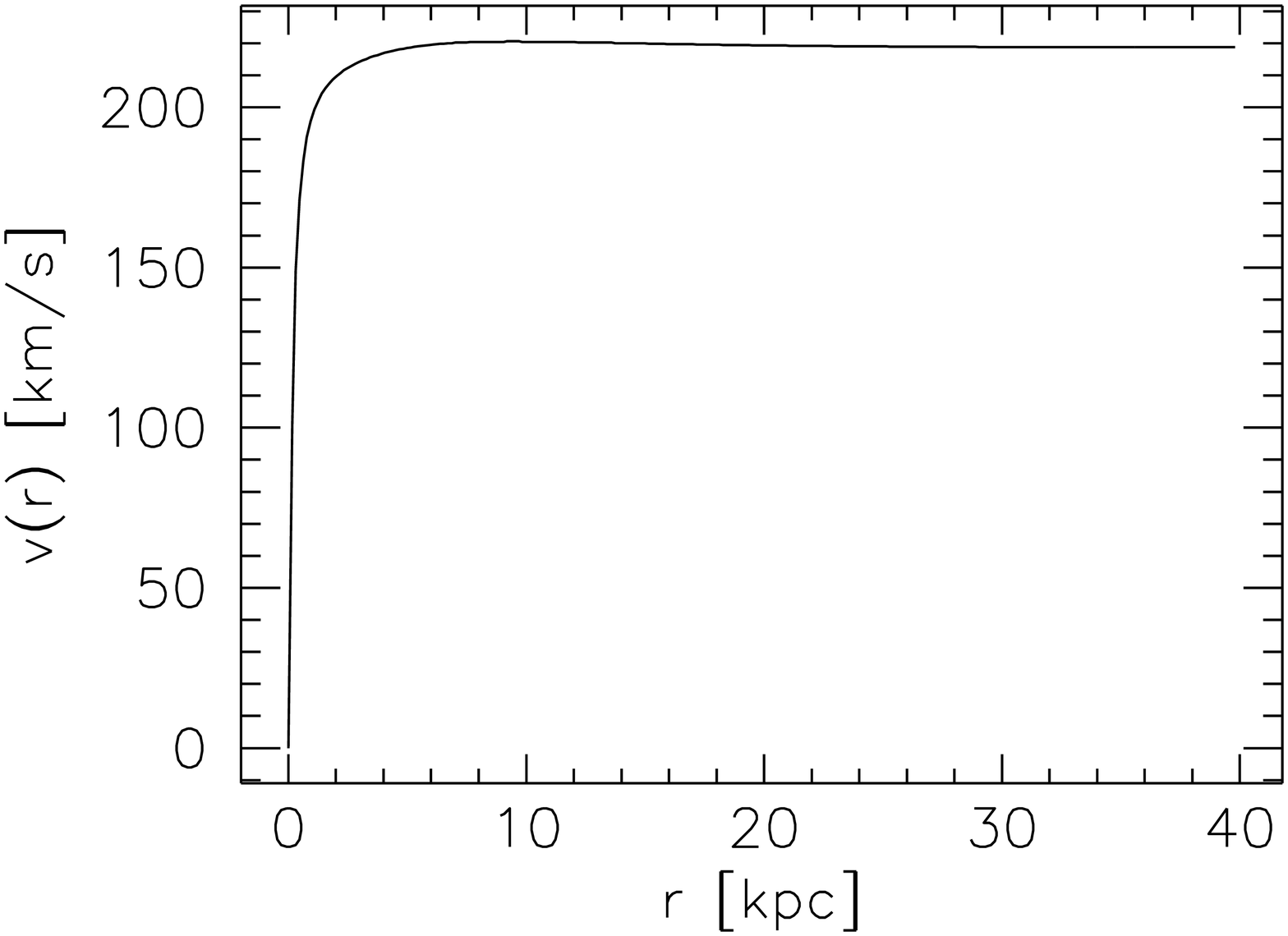}}
\caption{Rotation curve of a model consisting of a truncated Mestel
  disk embedded in an isothermal halo with $v_{\rm c}=220\,{\rm
  km\,s^{-1}}$. The curve steeply rises to $v_{\rm c}$ and then levels
  off, being flat to better than 5\% at large radii.}
\label{fig:1}
\end{figure}

To completely define the model, we need to specify five parameters,
namely the circular velocity $v_{\rm c}$, the two core radii $s_{\rm
d,h}$, the disk radius $r_{\rm d}$, and the disk oblateness $q_{\rm
3d}$. For reference, we define $v_{\rm c*}=220\,{\rm km\,s^{-1}}$,
$r_{\rm d*}=8\,h^{-1}\,{\rm kpc}$, and $q_{\rm 3d*}=0.03$. Requiring a
flat rotation curve, we must then use $s_{\rm h*}\approx0.72\,r_{\rm
d*}$ (Keeton \& Kochanek 1997). Finally, we choose $s_{\rm
d*}=0.2\,h^{-1}\,{\rm kpc}$. The rotation curve of this model is
plotted in Figure~\ref{fig:1}. As the figure shows, the rotation curve
is flat to better than 5\% beyond a radius of $\sim2\,h^{-1}\,{\rm
kpc}$.

\begin{figure}[ht]
  \centerline{\epsfxsize=0.5\hsize\epsffile{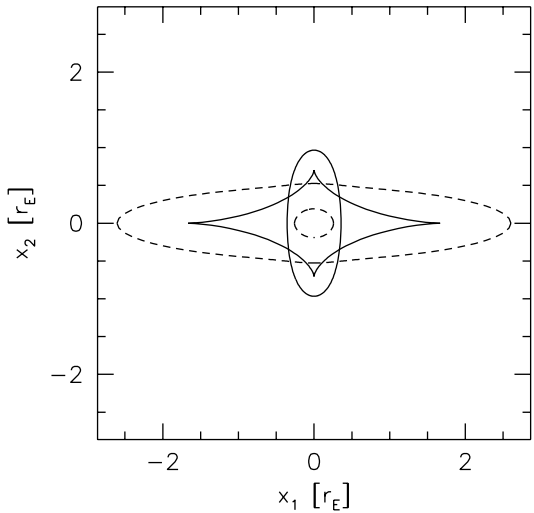}}
\caption{Critical curves (dashed lines) and caustics (solid lines) for
  a disk$+$halo lens. Since near edge-on configurations are preferred
  both geometrically and by the magnification bias, we show results
  for a disk inclination angle of $i=10^\circ$. The caustic
  corresponding to the outer critical curve shows the familiar astroid
  shape with four cusps connected by folds. The substantial
  contribution of the disk to the surface mass density stretches the
  caustic along the symmetry axis of the projected disk.}
\label{fig:2}
\end{figure}

To illustrate the lensing properties of the combined disk$+$halo model
described by equation~(\ref{eq:9}), we show in Figure~\ref{fig:2} the
caustics and critical curve configuration for a disk inclination angle
of $i=10^\circ$. The lens has two critical curves and caustics. The
inner caustic, which is the image of the outer critical curve, has the
familiar astroid shape with four cusps connected by folds.

\subsection{Neutral Hydrogen Disk}

Next, we address the effects of the gaseous component of the disk. To
evaluate the level of dust extinction and HI absorption, we assume
that the HI density, $n_{\rm H}$, follows a double-exponential
profile,
\begin{equation}
  n_{\rm H}(R,x_3) = n_{\rm H,0}\,\exp\left(-\frac{|x_3|}{H}\right)\,
  \exp\left(-\frac{R}{R_0}\right)\;.
\label{eq:10}
\end{equation}
The three parameters that define this profile are the central neutral
hydrogen density $n_{\rm H,0}$, the disk scale height $H$, and the
disk scale length $R_0$. In order to keep the number of free
parameters small in our model, we assume $H=q_3\,r_{\rm d}$ and
$R_0=r_{\rm d}$, so that the HI scale height is given by the short
axis of the oblate ellipsoid used to describe the lensing disk. We
also identify the HI scale radius with the truncation radius for the
lensing disk. The radial shape of the resulting radial HI profile is
similar to that observed in local disk galaxies (Broeils \& van
Woerden 1994). We choose $n_{\rm H,0}$ so that the face-on HI column
density, $N_0=2Hn_{\rm H,0}$, corresponds to the characteristic
observed value (Broeils \& van Woerden 1994), $N_0\approx
11.25\times10^{20}\,{\rm cm^{-2}}$. Hereafter, we use the notation
$N=10^{20}\,N_{20}\,{\rm cm^{-2}}$. Note that with these parameter
choices, the solar neighborhood values for the HI column density
($N_{\rm HI}=7.4\times10^{20}\,{\rm cm}^{-2}$; Kulkarni \& Heiles
1987) and HI scale height (a few hundred pc; Knapp 1987; Kuijken \&
Gilmore 1989) of the Milky Way disk are reproduced reasonably well.

\subsection{Extinction by Dust}

\begin{figure}[ht]
  \centerline{\epsfxsize=0.5\hsize\epsffile{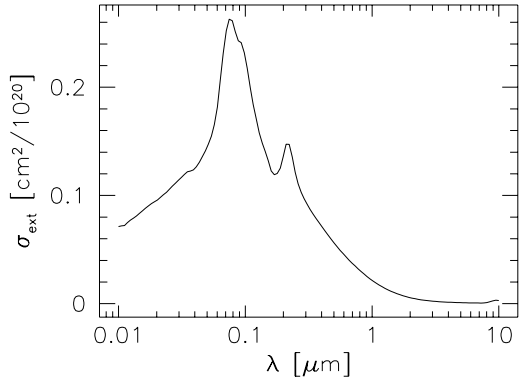}}
\caption{Sum of the extinction cross sections for scattering and
  absorption by graphites and silicates. The total cross section,
  $\sigma_{\rm ext}(\lambda)$, is given in units of ${\rm cm}^2$ per
  $10^{20}$ hydrogen atoms, assuming a gas-to-dust fraction of $1:100$
  by mass, following Draine \& Lee (1984).}
\label{fig:3}
\end{figure}

We assume that the distribution of dust follows equation~(\ref{eq:10})
with a constant ratio of dust to HI gas. We adopt the scattering and
absorption cross sections due to silicates and graphites derived by
Draine \& Lee (1984). The total extinction cross section is the sum of
the contributions from scattering and absorption, and the relative
proportion of these components is chosen so as to fit best the
observed Galactic extinction law. Figure~\ref{fig:3} shows the sum
$\sigma_{\rm ext}$ of the extinction cross sections from graphites and
silicates as a function of wavelength, assuming a dust-to-gas ratio of
$1:100$ by mass (Whittet 1992).

The dust optical depth is given by
\begin{equation}
  \tau_{\rm ext}(\lambda) = N\,\sigma_{\rm ext}(\lambda)\;,
\label{eq:11}
\end{equation}
so that the extinction in magnitudes is
\begin{equation}
  \Delta m(\lambda) = \left|2.5\,\log_{10}
  \{\exp[-N\sigma_{\rm ext}(\lambda)]\}
  \right| \approx 1.09\,N\,\sigma_{\rm ext}(\lambda)\;.
\label{eq:12}
\end{equation}

In quantifying the influence of dust, we restrict our attention to the
{\em observer's\/} Johnson $B$ band. In the rest frame of a lens at a
redshift $z_{\rm l}$, this band is centered on the wavelength
$\lambda_{\rm l}=0.435\,\mu{\rm m}\,(1+z_{\rm l})^{-1}$. For $z_{\rm
l}=0.3$, Figure~\ref{fig:3} yields $\sigma_{\rm
ext}N\approx0.084\,N_{20}$. Hence, for $N=10^{21}\,{\rm cm^{-2}}$,
$\Delta m\approx 0.9$ magnitudes in the $B$ band. Note that $N$ can
easily exceed this value for edge-on disks, which are favored both
geometrically and due to the magnification bias. The value of $N$ can
be as large as $N_0\,R/H=N_0/q_3\approx3.4\times10^{22}\,{\rm
cm^{-2}}$, yielding $\Delta m\approx31$ for a lens at $z_{\rm
l}=0.3$. This implies that the influence of dust on the imaging
properties of spiral lenses can be severe.

One would expect the dust-to-gas ratio of spiral disks to decline with
increasing redshift, in accordance with their metallicity
history. Indeed, Pei, Fall, \& Bechtold (1991) infer that the
dust-to-gas ratio in damped \lya absorbers at redshifts $2\la z\la3$
is only a tenth of the Milky-Way value, although with considerable
scatter (Fall \& Pei 1993). If we parameterize the dust content to
have a power-law dependence on redshift, $(1+z)^{-\delta}$, then a
reduction by a factor of $\sim10$ from the present time to
$z\approx2.3$ implies $\delta\approx2$. Equation~(\ref{eq:12}) is then
changed to
\begin{equation}
  \Delta m(\lambda,z_{\rm l}) \approx 1.09\,(1+z_{\rm l})^{-2}\,N_{20}\,
  \sigma_{\rm ext}\left(\frac{\lambda}{1+z_{\rm l}}\right)\;.
\label{eq:13}
\end{equation}
This dependence will be referred to as the ``dust evolution'' model in
the discussion that follows.

\subsection{Scalings of Galaxy Properties with Luminosity}

We assume that the lens population admits the Schechter luminosity
function at present with a number density per unit luminosity,
\begin{equation}
  {{\rm d}n(l)\over {\rm d}l}\,{\rm d}l = n_*\,l^\nu\,\exp(-l)\,{\rm d}l\;,
\label{eq:20}
\end{equation}
where $l\equiv L/L_*$ is the scaled galaxy luminosity. For spiral
galaxies, $n_*=1.5\times10^{-2}\,h^3\,{\rm Mpc}^{-3}$ and $\nu=-0.81$
(Marzke et al.\ 1994). The luminosity is related to the circular
velocity through the Tully-Fisher (1977) relation,
\begin{equation}
  \frac{v_{\rm c}}{v_{\rm c*}} = l^{1/\alpha}\;.
\label{eq:21}
\end{equation}
where $\alpha$ varies between $\sim2$ in the $B$ band and $\sim4$ in
the $H$ band (see review by Strauss \& Willick 1995, and references
therein).

Equations (\ref{eq:5}) and (\ref{eq:21}) imply that the Einstein
radius scales as $b\propto l^{2/\alpha}$. The computational effort
required later on is substantially reduced if the scale radii of the
disk$+$halo model change with $l$ in the same way as $b$ does. For
convenience, we therefore adopt the infrared value of $\alpha=4$ and
assume that the scale radii behave like\footnote{Since $r_{\rm
d}\propto l^{1/2}$ implies constant surface brightness, the chosen
scaling reflects Freeman's law (Freeman 1970; Holmberg 1975; Peterson,
Strom, \& Strom 1979; van der Kruit 1987; Lauberts \& Valentijn
1989). Thus, we effectively ignore the scatter in Freeman's law, and
the influence of low surface brightness (LSB) galaxies on the lensing
cross section (e.g., Bothun, Impey, \& McGaugh 1997 and references
therein). This omission is justified here because (i) the average
circular velocities of LSB galaxies is somewhat lower than that of
high surface brightness (HSB) galaxies, and the lensing cross section
is a sensitive function of the circular velocity, and (ii) LSB
galaxies are less compact than HSB galaxies and thus less efficient
lenses. For the purposes of strong lensing, it therefore seems safe to
neglect any contribution from LSB galaxies.} $r_{\rm d}\propto
l^{1/2}$. It is then sufficient to calculate all cross sections for
one reference luminosity only, and later scale the result to the
desired $l$ by changing the Einstein angle. If $\sigma_*$ is some
arbitrary cross section for the reference luminosity $L_*$, then the
average cross section for the entire population of spiral galaxies is
\begin{equation}
  \langle\sigma\rangle = n_*\,\int_0^\infty\,{\rm d}l\,
  \sigma_*\,l^{4/\alpha}\,l^\nu\,\exp(-l) =
  n_*\,\Gamma(1+\nu+4/\alpha)\,\sigma_*\;.
\label{eq:22}
\end{equation}

\subsection{Evolution of the Spiral Galaxy Population}
\label{subsec:2.5}

The simplest assumption about the evolution of the spiral population
is that they maintain a constant comoving number density,
$n(z)=n(0)(1+z)^3$, and constant scale radii. This is the {\em
no-evolution\/} model. Alternatively, Mo, Mao, \& White (1997)
recently suggested a model for the evolution of spirals, based on four
assumptions: (i) the disk mass is a fixed fraction $m_{\rm d}$ of the
halo mass; (ii) the disk angular momentum is a fixed fraction $j_{\rm
d}$ of the halo's angular momentum; (iii) the radial disk profile is
exponential, and the disk is centrifugally supported; and (iv) the
disk is dynamically stable. These assumptions yield a set of simple
and straightforward scaling relations, which we reproduce from Mo et
al. (1997).

We assume that the halo is a singular isothermal sphere with a radial
density profile,
\begin{equation}
  \rho = \frac{v_{\rm c}^2}{4\pi Gr^2}\;.
\label{eq:23}
\end{equation}
Taking the virial radius $r_{200}$ as the size of the halo, the halo
mass is $M=v_{\rm c}^2r_{200}/G$. Here $r_{200}$ is the radius within
which the average halo density is 200 times the {\em critical\/}
density of the Universe,
\begin{equation}
  \frac{3\,M}{4\pi\,r_{200}^3} = 200\times\frac{3H^2(z)}{8\pi G}\;,
\label{eq:24}
\end{equation}
and hence,
\begin{equation}
  r_{200}(z) = \frac{v_{\rm c}}{10\,H(z)} \approx
  220\,h^{-1}\,{\rm kpc}\,
  \left(\frac{v_{\rm c}}{220\,{\rm km\,s^{-1}}}\right)\,
  \left(\frac{H(z)}{H_0}\right)^{-1}\;,
\label{eq:25}
\end{equation}
where $H_0=100\,h\,{\rm km~s^{-1}~Mpc^{-1}}$ is the Hubble constant.
Similarly,
\begin{equation}
  M(z) = \frac{v_{\rm c}^3}{10\,G\,H(z)} \approx
  2.5\times10^{12}\,h^{-1}\,M_\odot\,
  \left(\frac{v_{\rm c}}{220\,{\rm km\,s^{-1}}}\right)^3\,
  \left(\frac{H(z)}{H_0}\right)^{-1}\;.
\label{eq:26}
\end{equation}
The disk radius $r_{\rm d}$ is related to the spin parameter of the
halo, $\lambda$ ($\equiv J\,|E|^{1/2}/GM^{5/2}$, where $J$ and $E$ are
the total angular momentum and energy of the halo, respectively), by
\begin{equation}
  r_{\rm d} = \frac{\lambda v_{\rm c}}{10\sqrt{2}\,H(z)}\,
  \left(\frac{j_{\rm d}}{m_{\rm d}}\right) \approx
  7.8\,h^{-1}\,{\rm kpc}\,\left(\frac{\lambda}{0.05}\right)\,
  \left(\frac{v_{\rm c}}{220\,{\rm km\,s^{-1}}}\right)\,
  \left(\frac{j_{\rm d}}{m_{\rm d}}\right)\,
  \left(\frac{H(z)}{H_0}\right)^{-1}\;,
\label{eq:27}
\end{equation}
and the central surface mass density of the disk is\footnote{Different
realizations of galaxy formation in different tidal environments could
result in disks of different sizes for the same halo properties, due
to variations in the spin parameter acquired by the baryons. As
discussed before, we ignore this scatter in the disk properties here.}
\begin{eqnarray}
  \Sigma_0 &=& \frac{10}{\pi}\,
  \frac{m_{\rm d}v_{\rm c}H(z)}{G\lambda^2}\,
  \left(\frac{m_{\rm d}}{j_{\rm d}}\right)^2 \nonumber\\
  &\approx&
  6.8\times10^{-2}\,h\,{\rm g\,cm^{-2}}\,
  \left(\frac{m_{\rm d}}{0.05}\right)\,
  \left(\frac{\lambda}{0.05}\right)^{-2}\,
  \left(\frac{v_{\rm c}}{220\,{\rm km\,s^{-1}}}\right)\,
  \left(\frac{m_{\rm d}}{j_{\rm d}}\right)^2\,
  \left(\frac{H(z)}{H_0}\right)\;.
\label{eq:28}
\end{eqnarray}

In summary, scale radii scale like $\propto v_{\rm c}\,H(z)^{-1}$, and
the surface mass density\footnote{We ignore the evolution of the HI
mass fraction of the disk due to star formation, as most of the star
formation occurred at redshifts $1.5\la z_{\rm l}\la2.5$ (Madau 1997),
while most of the lensing probability is contributed between $0.3\la
z_{\rm l}\la0.7$.} scales like $\propto v_{\rm c}\,H(z)$. Since the
characteristic values of $\lambda$, $j_{\rm d}$, and $m_{\rm d}$ are
expected to depend very weakly on redshift, we use these simple
proportionality relations with $v_{\rm c}$ and $H(z)$ in scaling the
properties of the local spiral population to higher redshifts. Since
$H(z)/H_0>1$ for $z>0$, scale radii {\em decrease\/} with increasing
$z$, while the surface mass density {\em increases\/} with increasing
$z$.

The evolution in the number density of spiral galaxies can be
expressed in terms of the Press-Schechter distribution function for
the mass $M$, given by equation~(\ref{eq:26}). We evolve the number
density of galaxies given by equations~(\ref{eq:20}), (\ref{eq:21}),
and (\ref{eq:26}) with the factor $n_{\rm PS}(M,z)/n_{\rm PS}(M,z=0)$,
where $n_{\rm PS}(M,z)$ is the Press-Schechter number density of halos
with mass $M$ at redshift $z$. Because of the inherent uncertainty in
modeling galaxy evolution, we present numerical results for three
models, assuming: (i) no evolution; (ii) evolution of scale radii
only; and (iii) evolution of galaxy number density and scale radii.

It is well known that the present-day Press-Schechter mass function
extends out to halo masses which are well beyond the galactic mass
scale (e.g., Navarro, Frenk, \& White 1995) in all viable models of
structure formation. This is due to the fact that the nonlinear mass
scale at present (which defines the exponential break in the
Press-Schechter mass function) corresponds to a much larger circular
velocity than $L_*$ in the Schechter function does, based on the
Tully-Fisher relation [cf. Eqs.~(\ref{eq:20}) and~(\ref{eq:21})]. This
implies that non-gravitational processes (such as inefficient cooling,
or expulsion of gas by supernovae) prevented disk formation inside
super-galactic halos at the present time. An evolutionary model for
galaxies based solely on the Press-Schechter approach is therefore
incomplete. The simplest interpretation of the discrepancy between the
Press-Schechter mass function and the local luminosity function of
galaxies is that massive galaxies with $L\ga L_*$ did not change their
dynamical properties since the redshift of galaxy formation
($z\sim2-4$) when the nonlinear mass scale was comparable to their
mass. Indeed, recent observations imply no significant evolution in
the population of massive galaxies out to redshifts $z\sim1$ (Ellis
1997, and references therein; but see Kauffmann, Charlot, \& White
1996). However, the same observations reveal many more dwarf galaxies
at high redshift than found locally. Since the lensing probability is
dominated by $L_*$ galaxies at $z\la0.7$, the no-evolution model might
be more appropriate for calculations of lensing by disk
galaxies. However, to bracket the other extreme of complete evolution
we show results also for the Press-Schechter prediction. Since we use
the Press-Schechter mass function only to correct the overall
normalization of the number density of spiral galaxies relative to its
present-day value, our approach should be less affected by the
incompleteness of the Press-Schechter treatment.

In all numerical calculations, we use the cosmological parameters
$\Omega_0=0.3$, $\Omega_\Lambda=0$, and $h=0.7$. For the
Press-Schechter mass function, we use the standard CDM power spectrum
(Bardeen et al. 1986) with the normalization $\sigma_{8h^{-1}{\rm
Mpc}}=1$. A cosmological model with these parameters reproduces the
local abundance of rich galaxy clusters (White, Efstathiou, \& Frenk
1993; Eke, Cole, \& Frenk 1996; Viana \& Liddle 1996), has the shape
parameter $\Omega_0\,h=0.21$ preferred by analyses of galaxy
clustering (Peacock \& Dodds 1994), but has a somewhat higher
normalization than derived from the COBE data (e.g. Ratra et
al. 1997). It also agrees with the observed abundance of giant
luminous arcs in galaxy clusters (Bartelmann et al. 1997).

\section{Lensing Cross-Sections}

Next, we proceed to calculate the magnification cross section of the
disk$+$halo lenses. This calculation must be done numerically. First,
we cover the source plane with a grid of source positions. Far from
the lens center, the resolution of this grid can be low, while close
to the caustic curves where the highest magnifications arise, the
resolution should be high. We therefore use an adaptive grid in the
source plane whose resolution increases towards the caustic
curves. Then, for each source position, all image positions need to be
found. For this purpose, we use the algorithm described by Schneider,
Ehlers, \& Falco (1992). Briefly, it is based on covering the image
plane with an uniform grid, which is then mapped back to the source
plane. All grid cells on the image plane whose mapping on the source
plane contains the source position, are taken to contain images of the
source. In regions of strong lensing, the parity of the cell might be
flipped when it is mapped into the source plane, i.e. corners of the
cell might be interchanged. One should therefore start with triangular
rather than rectangular cells in the image plane, because the image of
a triangle remains a convex figure whose interior is well-defined.
This can be easily achieved by splitting each rectangular cell along
one of its diagonals. Here again, we use an adaptive-grid approach in
order to achieve high resolution at a reasonable computational
cost. First, the image positions are searched on a coarse grid,
constraining the regions on the lens plane where the images are
located. Then, each of these regions is covered with a fine grid on
which the final image positions are localized. This way, all the
images corresponding to every segment on the source-plane grid are
identified.

The magnification of each image can then be computed from the total
lensing potential (\ref{eq:9}),
\begin{equation}
  \mu_{\rm GL} = \det{}^{-1}\left(\delta_{jk}-
  \frac{\partial^2\psi}{\partial x_j\partial x_k}\right)\;,
\label{eq:14}
\end{equation}
evaluated at each of the image positions $\vec x_i$. When dust
extinction is included, the effective magnification of an image at
position $\vec x_i$ is
\begin{equation}
  \mu(\vec x_i) = \mu_{\rm GL}(\vec x_i)\,
  \exp[-N(\vec x_i)\sigma_{\rm ext}]\;,
\label{eq:15a}
\end{equation}
where $N(\vec x_i)$ is the HI column density at the position of the
image. The net magnification is the sum of the moduli of the effective
magnifications for all of its $N$ images,
\begin{equation}
  \mu = \sum_{i=1}^{N}|\mu(\vec x_i)|\;.
\label{eq:15}
\end{equation}

\begin{figure}[ht]
  \centerline{\epsfxsize=0.5\hsize\epsffile{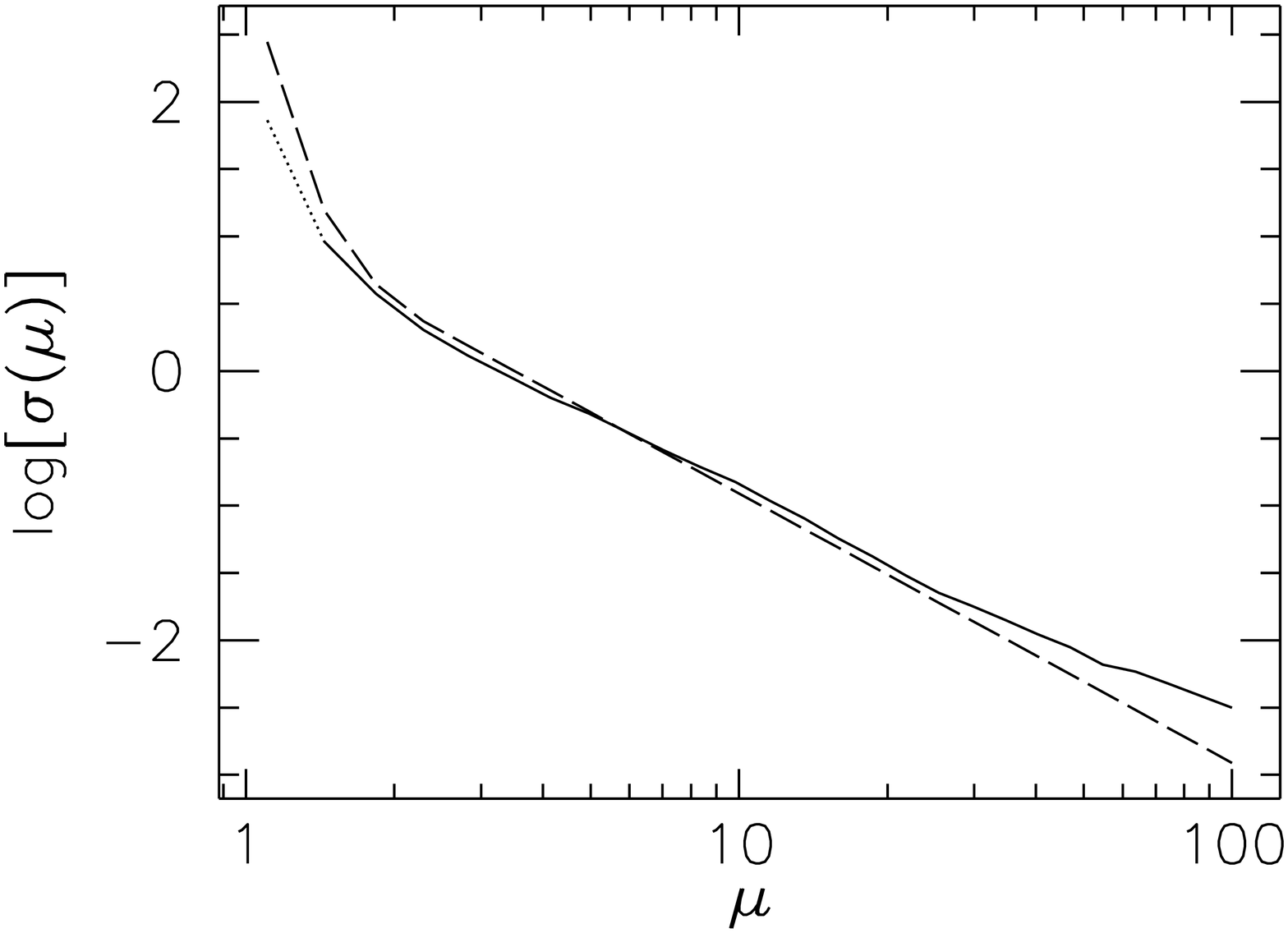}}
\caption{Inclination-averaged magnification cross section of a
  disk$+$halo lens (solid line) compared to that of a singular
  isothermal sphere (dashed line) with the same circular velocity. The
  source redshift is $z_{\rm s}=2$, and the lens redshift is $z_{\rm
  l}=0.3$. The cross sections are given in units of the Einstein disk
  area, $\pi b^2$. At small magnifications ($\mu\sim 1$) the cross
  section of the spiral lens becomes difficult to evaluate
  numerically, as indicated by the dotted section of the solid
  curve. The cross sections are almost identical, except for high
  magnifications, where the disk$+$halo model has a slightly larger
  cross section than the singular isothermal sphere.}
\label{fig:4}
\end{figure}

This approach yields maps of the source magnification in the source
plane. The magnification cross sections can be extracted from these
maps as the area in the source plane within which sources are
magnified by a factor $\ge\mu$. The magnification cross section
depends on the orientation of the disk relative to the
line-of-sight. Denoting the cross section for a magnification $\ge\mu$
and a disk inclination angle $i$ by $\sigma'(\mu,i)$, we obtain the
inclination-averaged cross section through the integral,
\begin{equation}
  \sigma(\mu) = \int_0^{\pi/2}\,{\rm d}i\,\cos(i)\,
  \sigma'(\mu,i)\;.
\label{eq:16}
\end{equation}
Obviously, $\sigma(\mu)$ depends on the cosmological distances
involved and on the other lens parameters. We suppress these
dependences here for brevity.

Figure~\ref{fig:4} shows that the inclination-averaged magnification
cross section of the disk$+$halo model is almost identical to that of
a singular isothermal sphere with the same circular velocity. Without
imposing further conditions, like a minimum image separation or a
maximum flux ratio between the images, the lensing statistics of
spiral lenses is well described by the simple spherical halo model. We
will demonstrate below what happens when further constraints are
imposed on the image properties.

In the limit of small magnifications ($\mu\to1$), the cross section
for the disk$+$halo model becomes numerically incomplete; this follows
from the fact that the cross-section increases rapidly as $\mu\to1$,
while the simulation is spatially bounded. This limit is irrelevant
for our purposes because we always impose further imaging
constraints. In particular, when we require the image separation and
flux ratio to be bounded to reasonable limits, the cross section is
confined to the multiple-imaging region, which is entirely contained
within our simulated section of the source plane.

\subsection{Imaging Probabilities}

Given the inclination-averaged cross sections, the probability for a
(point) source at redshift $z_{\rm s}$ to be imaged with magnification
$\ge\mu$ is obtained by the line-of-sight integral over the density of
lenses times their lensing cross section, $\sigma(\mu)$. For the
model without number-density evolution, this integral yields
\begin{equation}
  P'_{\rm GL}(\mu) = n_*\,b_*^2\,
  \Gamma(1+\nu+4/\alpha)\,\int_0^{z_{\rm s}}\,{\rm d}z\,
  (1+z)^3\,\left(\frac{D_{\rm ls}}{D_{\rm s}}\right)^2\,
  \left|\frac{c{\rm d}t}{{\rm d}z}\right|\,
  \sigma(\mu)\;,
\label{eq:17}
\end{equation}
where $|c{\rm d}t/{\rm d}z|$ is the proper-distance interval
corresponding to the redshift interval ${\rm d}z$. The factor
$n_*\,b_*^2\,\Gamma(1+\nu+4/\alpha)$ comes from integrating over the
luminosity distribution of the spiral galaxies [cf.\
Eq.~(\ref{eq:22})]. When evolution of the galaxy population is taken
into account as described in \S\ref{subsec:2.5}, the factor $(1+z)^3$
above is changed to $n_{\rm PS}(M_*,z)/n_{\rm PS}(M_*,0)\,(1+z)^3$,
where $M_*(z)$ is the mass of a galaxy with circular velocity $v_{\rm
c*}$ at the corresponding redshift.

Now let $|{\rm d}N_{\rm QSO}/{\rm d}S|(S){\rm d}S$ be the intrinsic
number density of quasars at redshift $z_{\rm s}$ with a flux between
$S$ and $S+{\rm d}S$. Accounting for magnification bias, the number of
lensed quasars is
\begin{equation}
  N'_{\rm QSO}(S) = \int_0^\infty\,{\rm d}S'\,
  P'_{\rm GL}\left(\frac{S}{S'}\right)\,
  \frac{{\rm d}N_{\rm QSO}}{{\rm d}S}(S')\;,
\label{eq:18}
\end{equation}
and hence the probability for a quasar at redshift $z_{\rm s}$ to be
detected with flux $>S$ is
\begin{equation}
  P_{\rm GL}(S) = \frac{1}{N_{\rm QSO}(S)}\,
  \int_0^\infty\,{\rm d}S'\,P'_{\rm GL}\left(\frac{S}{S'}\right)\,
  \frac{{\rm d}N_{\rm QSO}}{{\rm d}S}(S')\;.
\label{eq:19}
\end{equation}
We approximate the observed quasar number counts in the $B$-band by a
broken power-law,
\begin{equation}
  \frac{{\rm d}N_{\rm QSO}}{{\rm d}S} = \left\{
  \begin{array}{ll}
  (S/S_0)^a & \hbox{for}\quad S\le S_0 \\
  (S/S_0)^b & \hbox{for}\quad S > S_0 \\
  \end{array}\right.\;,
\label{eq:29}
\end{equation}
where $S_0$ corresponds to $B_{\rm QSO}\approx 19$. For quasar
redshifts $z_s\sim 2$, $a=-1.64$ and $b=-3.52$ (e.g. Pei 1995).

\begin{figure}[ht]
  \centerline{\epsfxsize=0.5\hsize\epsffile{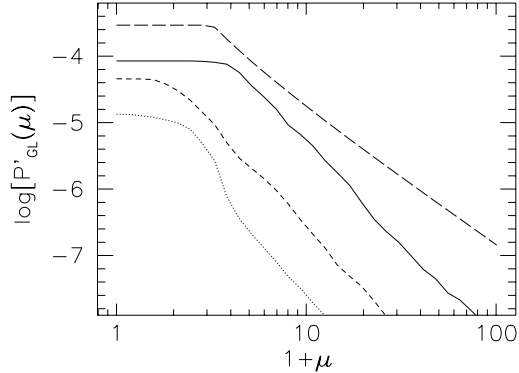}}
\caption{Probability $P'_{\rm GL}(\mu)$ for a quasar at a redshift
  $z_{\rm s}=2$ to be multiply imaged with magnification $>\mu$, image
  separation $>0.3''$, and flux ratio of the images $<20$. The curves
  are for the disk$+$halo model without dust (solid line), with
  non-evolving dust (dotted line), and with evolving dust
  (short-dashed line). We compare these results to a mass model of a
  singular isothermal sphere with no dust and the same asymptotic
  circular velocity as the disk$+$halo model (long-dashed line). These
  curves do not include evolution of scale lengths or number density
  of the lens population. Here and in all the following figures,
  results for the singular isothermal sphere model are calculated
  without dust extinction.}
\label{fig:5}
\end{figure}

\begin{figure}[ht]
  \centerline{\epsfxsize=0.7\hsize\epsffile{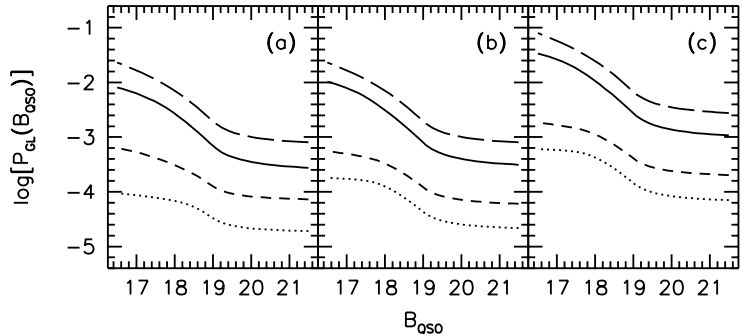}}
\caption{Probability $P_{\rm GL}(B_{\rm QSO})$ for a quasar to be
  multiply imaged with image separation $>0.3''$ and flux ratio $<20$,
  as a function of $B_{\rm QSO}$. The mass model includes no-evolution
  in panel (a), evolution of disk scale lengths in panel (b), and
  evolution of scale lengths and number density in panel (c). The four
  curves per panel are for the disk model without dust (solid curve),
  with non-evolving dust (dotted curve), and with evolving dust
  (short-dashed curve). The long-dashed curve shows results for the
  singular isothermal sphere model.}
\label{fig:6}
\end{figure}

Figure~\ref{fig:5} shows $P'_{\rm GL}(\mu)$, and Figure~\ref{fig:6}
shows $P_{\rm GL}(S)$ for the no-evolution galaxy model. These figures
include curves that illustrate the influence of non-evolving and
evolving dust. All curves in Figure~\ref{fig:5} were calculated under
the additional constraints that the the two brightest images be
separated by $\ge0.3''$, and that their flux ratio be $\le20$. The
solid and long-dashed curves in Figure~\ref{fig:5} are for the
disk$+$halo model and for the singular isothermal sphere,
respectively. Evidently, the disk$+$halo model is less efficient at
producing multiple images with the specified properties than the
singular isothermal sphere is, despite the fact that the total
magnification cross sections of the two models are close to each other
(Fig.~\ref{fig:4}). The reason for this difference is that the images
produced by the spiral lens model are typically closer to each other
than those of the singular isothermal sphere, and are occasionally
below the $0.3''$ threshold. This occurs because a significant
fraction of the lensing cross section is contributed by the disk in
nearly edge-on orientations. The image separation on either side of
the disk is then smaller than the Einstein diameter of the
corresponding singular isothermal sphere.

The imposed constraints on the minimum image separation and maximum
flux ratio approximately reflect characteristic thresholds for
detecting a multiply imaged system through space-based observations
with finite resolution and dynamical range (cf. Kochanek
1993). Figure~\ref{fig:5} demonstrates that these additional
constraints reduce the detection efficiency of spiral lenses
considerably compared to the singular isothermal sphere model. The
figure also quantifies the severe effect that dust extinction has on
the detection efficiency. Besides suppressing the detection
probability by one or two orders of magnitude for evolving and
non-evolving dust, respectively, the dust extinction allows for values
of $\mu=(S_{\rm observed}/S_{\rm intrinsic})$ smaller than unity,
which are otherwise unphysical.

Figure~\ref{fig:6} then shows the probability $P_{\rm GL}(S)$ for
quasars at redshift $z_{\rm s}=2$ to be lensed with image separation
$\ge0.3''$ and flux ratio $\le20$ as a function of the quasar $B$
magnitude. Again, the dotted and short-dashed curves are calculated
including dust extinction, and the solid and long-dashed curves are
without dust for the disk$+$halo model and for the singular isothermal
sphere model, respectively. The three panels in the figure are for
different levels of cosmological evolution as detailed in the figure
caption. The figure shows that bright quasars, with $B_{\rm
QSO}\le19$, are much more likely to be multiply imaged than faint
quasars, due to the magnification bias. At $B_{\rm QSO}\sim18$, the
multiple-imaging probability is about a factor of five higher than for
$B_{\rm QSO}\sim19$. As already shown in Figure~\ref{fig:5}, dust
extinction severely reduces the imaging probability, and it also leads
to a weaker rise in the lens detection probability with increasing
quasar brightness.

\subsection{Image Statistics}

\begin{figure}[ht]
  \centerline{\epsfxsize=0.7\hsize\epsffile{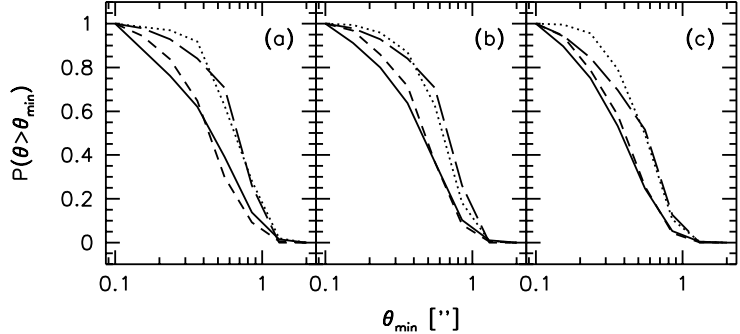}}
\caption{Cumulative distribution of multiple images with separation
  $\theta>\theta_{\rm min}$, as a function of $\theta_{\rm min}$, for
  a constant maximum flux ratio $r_{\rm max}=20$ and quasar magnitude
  $B_{\rm QSO}=18$. The disk model includes no-evolution in panel (a),
  evolution of scale lengths in panel (b), and evolution of scale
  lengths and disk number density in panel (c). The three curves per
  panel are without dust extinction (solid line), with extinction by
  non-evolving (dotted) and evolving (short-dashed) dust. The
  long-dashed curve describes the singular isothermal sphere model
  with no dust.}
\label{fig:7}
\end{figure}

In order to appreciate the significance of the selection effects in
identifying spiral lenses, it is instructive to examine the
distribution of image separations and flux ratios. Figure~\ref{fig:7}
shows the cumulative distribution of multiple images as a function of
the minimum image separation $\theta_{\rm min}$. The distribution is
arbitrarily normalized to unity at $\theta_{\rm min}=0.1''$. As
before, the three panels in the figure are for disks with no-evolution
(panel a), with evolution of scale radii only (panel b), and with
evolution of scale radii and galaxy number density (panel c). The
three curves per panel are for models without dust (solid line), with
non-evolving dust (dotted), and with evolving dust (short-dashed). The
long-dashed curve in each panel shows the singular isothermal sphere
model without dust for comparison.

Figure~\ref{fig:7} implies that the median image separation is larger
for the singular isothermal sphere than for the disk$+$halo
model. Images produced by the spiral lens model are on average closer
to each other by $\sim0.2''$ than expected from the singular
isothermal sphere model. This is because multiple images in the
dust$+$halo model are predominantly produced by the disk rather than
by the halo, with a preference for nearly edge-on disk orientations.

Dust increases the average image separation because images with small
separations occur close to the disk, where their flux is heavily
suppressed by extinction. Disk evolution reduces the average image
separation slightly, because the disks get more compact with
increasing redshift. Number density evolution according to the
Press-Schechter model increases the abundance of field spirals at
redshifts $z\sim1-2$ (since some of these galaxies have subsequently
merged and disappeared by now), and reduces the image separation
further because the lenses are on average further away.

\begin{figure}[ht]
  \centerline{\epsfxsize=0.7\hsize\epsffile{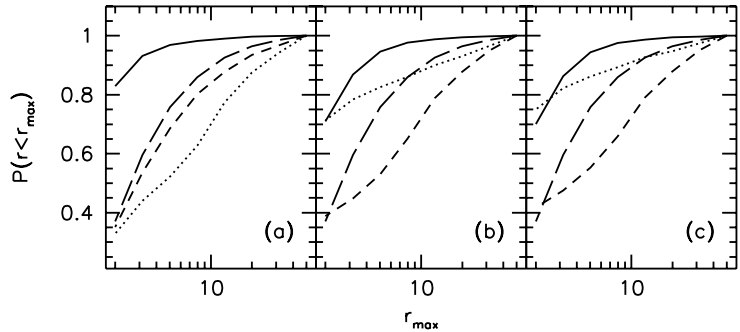}}
\caption{Cumulative distribution of multiple images with a $B$ band
  flux ratio $r<r_{\rm max}$, as a function of $r_{\rm max}$, for a
  constant $\theta_{\rm min}=0.3''$ and $B_{\rm QSO}=18$. The panels
  and the curves have the same meaning as in the previous figure. }
\label{fig:8}
\end{figure}

Figure~\ref{fig:8} shows the cumulative distribution of images as a
function of the maximum flux ratio $r_{\rm max}$, arbitrarily
normalized to unity at $r_{\rm max}=50$. The notation is the same as
in Figure~\ref{fig:7}. Generally, the curves for the disk$+$halo model
without dust are the flattest, showing that the images are usually of
comparable brightness. When dust is included, the average flux ratio
increases due to the strong extinction gradient around the disk.

\section{Effects of the Gaseous Disk}

\subsection{Impact of Dust on Lensing Statistics}

\begin{figure}[ht]
  \centerline{\epsfxsize=0.7\hsize\epsffile{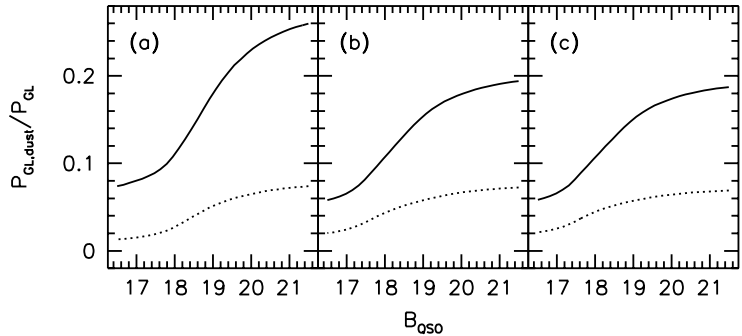}}
\caption{Reduction in the lens detection probability due to dust
  extinction in the (observer's) $B$ band. We assume $\theta_{\rm
  min}=0.3''$ and $r_{\rm max}=20$. The curves show the lensing
  probability including dust divided by the lensing probability
  without dust, as functions of the quasars' $B$ magnitude. The disk
  model has no-evolution in panel (a), evolution of scale lengths in
  panel (b), and evolution of scale lengths and disk number density in
  panel (c). The two curves per panel are for non-evolving (dotted
  line) and evolving (solid line) dust.}
\label{fig:9}
\end{figure}

To further examine the influence of dust on the detectability of
lensed quasars, we show in Figure~\ref{fig:9} the ratio between the
lensing probability $P_{\rm GL}(B_{\rm QSO})$ with and without
dust. The three panels are again for a galaxy model with no evolution
(panel a), with length scale evolution only (panel lb), and with
length-scale plus number-density evolution (panel c). The two curves
per panel are for non-evolving and evolving dust (dotted and solid
lines, respectively).

Figure~\ref{fig:9} shows that the majority of all quasars which are
lensed by spirals are undetectable in the $B$ band. The deficit caused
by dust increases for brighter quasars as a result of the
magnification bias. At $B_{\rm QSO}\approx18$, only about 10\% of all
lensed quasars are observable, quite independent of the degree of
evolution that is included in the calculation. If dust does not evolve
with redshift, the deficit of lensed quasars is more severe. For faint
quasars, $B_{\rm QSO}\ga20$, the reduction in the lens detection
probability is least severe in the no-evolution model. This is mainly
because when disk evolution is included, the surface density of the
disk increases with $z$, leading to more extensive dust extinction
than the no-evolution model predicts.

Figure~\ref{fig:9} implies that the fraction of quasars lensed by
spiral galaxies in radio surveys should be higher by about an order of
magnitude relative to that found in optical surveys.

\subsection{Statistics of Damped \lya Absorption}

\begin{figure}[ht]
  \centerline{\epsfxsize=0.7\hsize\epsffile{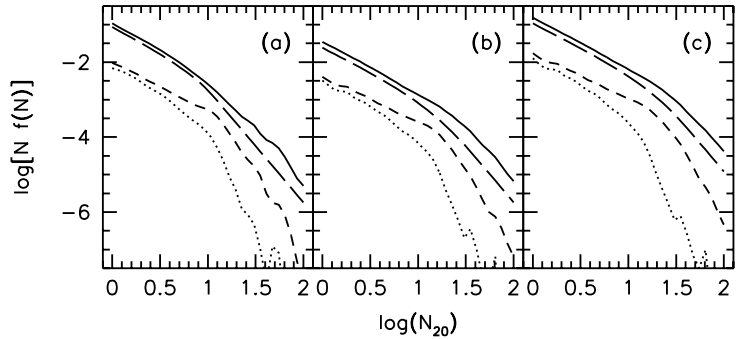}}
\caption{The HI column-density distribution, $Nf(N)$, for lensed
  quasars with $B=18$ at $z_{\rm s}=2$ as a function of $N_{20}\equiv
  (N/10^{20}~{\rm cm^{-2}})$. The disk model includes no-evolution in
  panel (a), evolution of scale lengths in panel (b), and evolution of
  scale lengths and number density in panel (c). The three curves were
  calculated without dust, with non-evolving dust, and with evolving
  dust (solid, dotted, and short-dashed curves, respectively). For
  reference, the long-dashed curve shows the result without lensing or
  dust extinction.}
\label{fig:10}
\end{figure}

Based on the neutral hydrogen column density at the position of each
image, we can investigate the influence of lensing and dust on the
statistics of damped Lyman-$\alpha$ absorption by spirals. Let $P_{\rm
GL}(S,N)$ be the probability to observe a quasar with a flux $\ge S$
which is imaged by the population of spirals and shows an HI column
density $>N$ in its spectrum. When there is more than one image, we
take $N$ to be the HI column density in the brightest image, i.e. that
which dominates the absorption trough. As shown by Bartelmann \& Loeb
(1996), the observed column-density distribution of neutral hydrogen
is then given by
\begin{equation}
  f(N) = \frac{c}{H_0}\,\frac{1}{\Delta X}
  \left|\frac{\partial P_{\rm GL}(S,N)}{\partial N}\right|\;,
\label{eq:30}
\end{equation}
where $\Delta X$ is the absorption distance scanned by the damped \lya
absorbers (DLAs) in the survey (Bahcall \& Peebles 1969). Examples for
the distribution of $Nf(N)$ are plotted in Figue~\ref{fig:10}.

\begin{figure}[ht]
  \centerline{\epsfxsize=0.7\hsize\epsffile{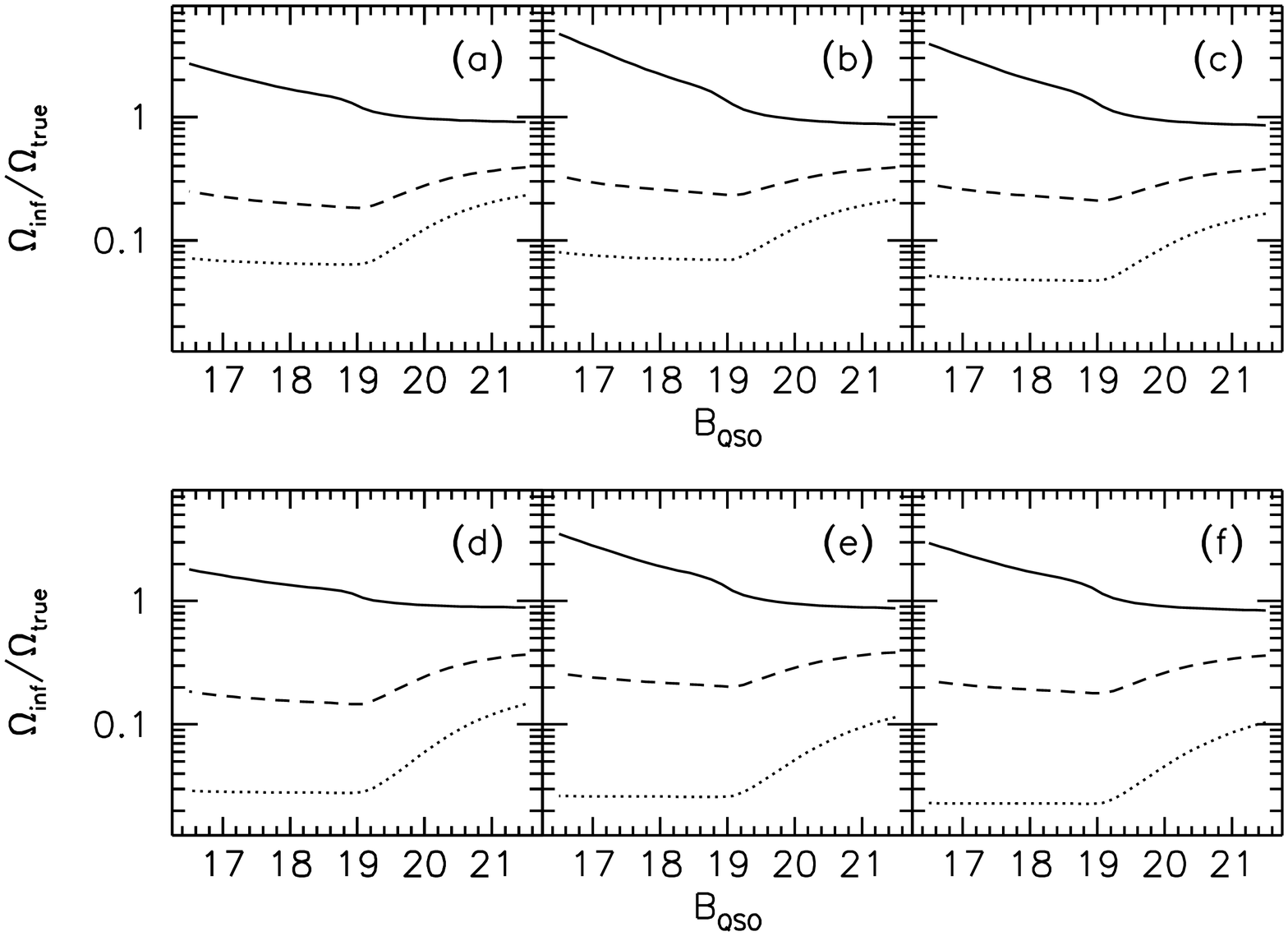}}
\caption{The ratio between the inferred and the true values of the
  density parameter of neutral hydrogen as a function of the quasar
  $B$ magnitude. For the upper three panels, the redshift range for
  the absorbers is $0\le z\le z_{\rm s}=2$, while for the lower panels
  it is restricted to $1\le z\le z_{\rm s}$. The disk model includes
  no-evolution in panels (a) and (d), evolution of scale lengths in
  panels (b) and (e), and evolution of scale lengths and number
  density in panels (c) and (f). The three curves were calculated
  without dust, with non-evolving dust, and with evolving dust (solid,
  dotted, and dashed curves, respectively).}
\label{fig:11}
\end{figure}

The inferred cosmological density parameter in neutral hydrogen,
$\Omega_{\rm HI}$, is given by
\begin{equation}
  \Omega_{\rm HI} = \frac{H_0}{c}\,\frac{\bar m}{\rho_{\rm cr,0}}\,
  \int_0^\infty\,{\rm d}N\,N\,f(N)\;,
\label{eq:31}
\end{equation}
where $\rho_{\rm cr,0}$ is the present-day critical density of the
Universe, and $\bar m$ is the mean molecular mass. Because of the
magnification bias, $f(N)$ depends on the quasar magnitude $B_{\rm
QSO}$, and so does the inferred $\Omega_{\rm HI}$. Figure~\ref{fig:11}
shows the ratio between the inferred and true values of $\Omega_{\rm
HI}$ for spiral galaxies. The different panels and line-types are the
same as those in previous figures. Similarly to $P_{\rm GL}$, $f(N)$
and $\Omega_{\rm HI}$ depend on the redshift range of the absorbers
[cf. Eq.~(\ref{eq:17})]. We use the full redshift range, $0\le z\le
z_{\rm s}=2$ for Figure~\ref{fig:10} and for the upper panels of
Figure~\ref{fig:11}, and the high-redshift range $1\le z\le z_{\rm s}$
for the lower panels of Figure~\ref{fig:11}.

Models without dust cause $\Omega_{\rm HI}$ to be overestimated by
factors of up to a few in samples of bright quasars. The magnification
bias brings into view quasars that were otherwise too faint to be
detected. The lensed quasar images occur close to the lens center,
where the HI column density is higher than average. However, when dust
is included, the net effect is reversed. Now those quasars whose
images are close to the disk are obscured, and the number of quasar
spectra with high HI column density is lower. For evolving dust and
$B_{\rm QSO}\approx18$, only about 20\%--30\% of the neutral hydrogen
is detected.

\section{Conclusions}

We investigated the lensing effects of spiral galaxies, modelled as
maximal truncated Mestel disks embedded in spherical halos. This model
makes the disk as massive as possible. Our results therefore represent
the opposite extreme to the simple isothermal sphere models.
Intermediate models with less massive disks, would shift these results
towards those obtained from the spherical model.

Although a disk$+$halo configuration yields an inclination-averaged
lensing cross section similar to that of the spherically symmetric
mass distribution (Fig.~\ref{fig:4}), its typical image separations
are considerably smaller (Fig.~\ref{fig:7}). A substantial
contribution to this cross section comes from lensing by nearly
edge-on configurations of the disk, in which a pair of images
straddles the disk. Because the images are often close to the disk,
they suffer strong extinction by dust. The extinction lowers the
detection efficiency of spiral lenses in the optical band by an order
of magnitude. Moreover, disk lenses are often characterized by small
image separation (Fig.~\ref{fig:7}) and large differential extinction
(Fig.~\ref{fig:8}). As a result, the selection effects imposed by the
finite angular resolution and dynamic range of observations set limits
on the minimum image separation and their maximum flux ratio, and
lower substantially the probability for observing spiral lenses (cf.
Fig.~\ref{fig:9}).

The spiral-lens system B~0218$+$357 shows compelling evidence for
strong extinction. O'Dea et al. (1992) find that the quasar spectrum
is red; Wiklind \& Combes (1995) and Menten \& Reid (1996) find strong
molecular lines; and Grundahl \& Hjorth (1995) find that image A is
much fainter than image B in the optical in contrast to the radio
observations, arguing for a substantial extinction of image A.
Jaunsen \& Hjorth (1997) have recently argued for the existence of
disk lensing and associated extinction also in the lens system
B~1600$+$434. More generally, Malhotra, Rhoads, \& Turner (1997) argue
for a systematic reddening of lensed quasars relative to the rest of
the quasar population.

The optical depth to gravitational lensing of quasars was shown to be
a very effective tool for setting constraints on the cosmological
constant (Kochanek 1996, and references therein). In singular
isothermal models, $\sim80\%-90\%$ of the lensing probability is
contributed by elliptical galaxies. However, if a significant fraction
of the present-day ellipticals were in the form of spiral building
blocks at $z\sim 1$ (as argued by Kauffmann, Charlot, \& White 1996),
then the depletion of spiral lenses due to dust extinction and
selection effects could weaken the current lensing constraints on the
cosmological constant. This issue was addressed by Mao \& Kochanek
(1994) and Rix et al. (1994). They found that estimates of the
cosmological constant from the statistics of strong lensing can be
significantly changed only when the elliptical galaxies seen today
evolved dramatically below a redshift $z\sim1$.

In this context, and in view of ongoing lensing surveys, it is
instructive to examine which fraction of the total lensing optical
depth is contributed by spirals rather than by ellipticals. To
calculate this, we assume that the total galaxy population today is
composed of 25\% ellipticals and 75\% spirals. We model spirals as
described above, and ellipticals as singular isothermal spheres with a
characteristic velocity dispersion of $\sigma_{v*}=v_{\rm
c*}/\sqrt{2}=220\,{\rm km\,s^{-1}}$, following the Schechter
luminosity function (\ref{eq:20}) and the Faber-Jackson (1976)
relation. For simplicity, we use the same Schechter-function
parameters as for the spirals. We consider two scenarios, one in which
the number density of ellipticals stays constant with redshift, and
another in which ellipticals are assembled by merging of spirals. In
the latter scenario, we assume that the number density of ellipticals
changes with redshift as a power-law, $n_{\rm E}(z) = n_{\rm
E}(0)\,(1+z)^\epsilon$, with $\epsilon\approx-1.6$, chosen such that
$\sim2/3$ of the ellipticals are in the form of spiral bulding blocks
at $z\sim1$ (Kauffmann et al. 1996). We further assume that mergers
conserve mass. In both scenarios, we can then compute the lensing
optical depths for both spirals and ellipticals $P_{\rm GL;E,S}'$ from
equation~(\ref{eq:17}) and the corresponding imaging probability
$P_{\rm GL;E,S}$ from equation~(\ref{eq:19}) (under the constraints
that the images be separated by $\ge0.3''$ and have a flux ratio
$\le20$).

\begin{figure}[ht]
  \centerline{\epsfxsize=0.7\hsize\epsffile{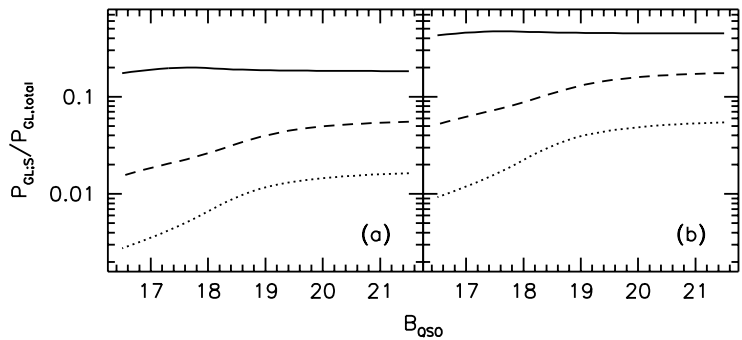}}
\caption{Fraction of the total multiple-imaging probability
  contributed by spiral galaxies. In panel (a) the elliptical
  population is assumed to have constant comoving number density, and
  in panel (b) the number density of ellipticals evolves as a power
  law of redshift so that $2/3$ of the ellipticals are in the form of
  spiral bulding blocks at $z=1$. The three curves per panel are for
  spirals without dust (solid line), with non-evolving dust (dotted
  line), and with evolving dust (dashed line). Apart from merging to
  ellipticals, any other evolution of the spiral population was
  ignored.}
\label{fig:12}
\end{figure}

Figure~\ref{fig:12} shows the fraction of the total lensing
probability contributed by spirals, namely $({P_{\rm GL,S}}/{P_{\rm
GL,total}}) = {P_{\rm GL,S}}/({P_{\rm GL,E}+P_{\rm GL,S}})$. Panel (a)
shows results for ellipticals with constant comoving number density,
while panel (b) examines the scenario in which ellipticals merge out
of spirals. The three curves per panel are for spirals without dust,
with non-evolving dust, and with evolving dust (solid, dotted, and
dashed lines, respectively). The no-dust results are valid for
radio-selected lenses. The solid line in panel (a) reproduces the
familiar result that dust-free spirals contribute at most $\sim20\%$
of the multiple-imaging probability, even for the small
image-separation cutoff of $\theta_{\rm min}=0.3''$; this fraction is
lower for higher $\theta_{\rm min}$. In the presence of dust, the
fraction contributed by spirals is considerably smaller. In the merger
scenario, the spiral lens fraction rises to $\sim50\%$ without dust,
and $\sim9\%$ with evolving dust, for quasars with $B\sim18$. Thus, up
to a half of the lenses found in radio surveys (which are sensitive to
small image separations) might be imaged by spirals.

Figure~\ref{fig:9} implies that radio surveys should be about a factor
of 5--10 more efficient at detecting spiral lenses than optical
surveys. The recent CLASS and JVAS radio surveys indeed provide
preliminary hints of a more substantial population of spiral lenses
than found in optical surveys (cf. Table~1 in Browne et al. 1997; and
Jackson, Nair, \& Browne 1997; but see Fassnacht \& Cohen 1997). To
date, the combined CLASS and JVAS samples encompass more than $10^4$
flat-spectrum radio sources, 11 of which have been identified as lens
systems. A substantial fraction of those has been classified as being
lensed by spirals or S0 galaxies (cf. Myers 1997; Williams \&
Schechter 1997). Because of their small image separation and simple
geometry, lenses containing edge-on disks might offer a unique
opportunity for estimating the masses of galactic disks at high
redshifts, and also for constraining the Hubble constant based on the
time-delay between the flux variations in their images. Indeed, the
time delay in B~0218 has been measured to $12\pm3$ days, translating
to a Hubble constant of $H_0\sim60\,{\rm km\,s^{-1}\,Mpc^{-1}}$
(Corbett et al. 1996), and B~1608$+$656 offers another promising lens
system for this purpose (Myers et al. 1995; Fassnacht et al. 1996).

Finally, we find that dust dominates over the magnification bias due
to lensing in modifying the statistics of damped \lya absorption by
spirals (Fig.~\ref{fig:11}). If most of the damped absorption systems
are spirals, the inferred value of the cosmological density parameter
in HI at $z\la1$ could be underestimated by a factor of a few.

The above results were obtained under the assumption that all spirals
possess the same dust-to-gas ratio throughout their disks at any given
redshift. If the scatter in the dust content of different galaxies is
large or if the dust distribution is patchy, then the quantitative
impact of dust on the lensing statistics would be altered. In
addition, if spiral disks at $z\la1$ are less massive than our {\em
maximal disk\/} model assumes, then the lensing cross section of the
disk and its related extinction signature would be reduced.

%JO \acknowledgments
\section*{Acknowledgements} %PP

We thank Chuck Keeton, Chris Kochanek, Shude Mao, Ed Turner, and Peter
Schneider for useful discussions. AL was supported in part by the NASA
ATP grant NAG5-3085 and the Harvard Milton fund. MB was supported in
part by the Sonderforschungsbereich 375 of the Deutsche
Forschungsgemeinschaft.

\end{document}